\PassOptionsToPackage{hyperfootnotes=false}{hyperref}
\documentclass[fleqn,usenatbib]{mnras}

\usepackage{amsmath,amssymb}
\usepackage{graphicx}
\usepackage{bm}
\usepackage[dvipsnames]{xcolor}
\usepackage{hyperref}
\usepackage{newtxtext,newtxmath}
\newcommand{\dd}{\mathrm{d}}

\makeatletter
\def\caption{%
  \ifx\@captype\@undefined
    \@latex@error{\noexpand\caption outside float}\@ehd
    \expandafter\@gobble
  \else
    \H@refstepcounter\@captype
    \let\Hy@tempa\@caption
    \expandafter\@firstofone
  \fi
  {\@dblarg{\Hy@tempa\@captype}}%
}
\makeatother

\title[Boundary-supported radial layering]{Boundary-supported radial layering in Hoag-like ring galaxies}

\author[C. Sun, Y. Xu, and J. Zhang]{
Chenxi Sun$^{1}$\thanks{E-mail: scx@pku.edu.cn.},
Yue Xu$^{2,3}$\thanks{Chenxi Sun and Yue Xu contributed equally to this work; their order is alphabetical and does not indicate priority.},
and Jianwei Zhang$^{4}$\thanks{E-mail: james@pku.edu.cn}
\\
$^{1}$Beijing Key Laboratory of Quantum Sensing and Precision Measurement, and\\
\hspace{1.1em}Center for Quantum Information Technology, and Institute of Quantum Electronics, Peking University, Beijing 100871, China.\\
$^{2}$Department of Physics, School of Physics and Materials Science, Nanchang University, Nanchang 330031, China\\
$^{3}$Henan Institute for Drug and Medical Device Control, Zhengzhou 450018, China\\
$^{4}$School of Physics, Peking University, Beijing 100871, China
}

\date{Accepted XXX. Received YYY; in original form ZZZ}
\pubyear{2026}

\begin{document}
\label{firstpage}
\pagerange{\pageref{firstpage}--\pageref{lastpage}}
\maketitle

\begin{abstract}
Clean Hoag-like ring galaxies are often characterized by an old compact
central component, a depleted gap, and a detached outer ring.  We
identify a boundary-supported radial-layering mechanism in a
shell-deformed Kepler control model.  A compact inner boundary supplies
the core state, while a localized effective shell deformation,
interpreted as the reduced imprint of externally supplied material
settled near a finite circularization radius, needs to create only an
internal maximum and a subsequent outer minimum.  These act as the gap
barrier and ring-supporting well.  The onset of this structure is
organized by a saddle-node threshold of the critical-point equation.  In
a $10^4$-point Monte Carlo scan, shell-localized boundary-supported
candidates occupy finite parameter volume under the adopted priors, and
none of the localized candidates contains an ordered interior
minimum--maximum--minimum subsequence.  The same branch gives a
scale-free gap-to-ring interval overlapping representative ratios for
Hoag's Object, UGC 4599, and PGC 1000714, but not for the
environmentally processed comparison object JO171.
\end{abstract}


\section{Problem and positioning.}
Hoag-like ring galaxies are distinguished not simply by the presence of
a ring, but by a clean radial separation: a compact central body, an
underdense gap, and a detached outer annulus.  This structure is
difficult to reduce to the standard ring-formation channels.  Nearly
head-on collisions generate expanding density waves
\citep{LyndsToomre1976,AppletonStruckMarcell1996}, while bar or oval
perturbations organize gas near resonances in disk galaxies
\citep{ButaCombes1996}.  The cleanest Hoag-type systems, however, often
show spheroidal or nearly round cores, weak or absent bar signatures, and
evidence for externally supplied gas or extended H~{\sc i} reservoirs
\citep{Schweizer1987,Finkelman2011Hoag,FinkelmanBrosch2011UGC4599,
MutluPakdil2017PGC1000714,Moretti2018JO171}.

These observations motivate a more specific structural question.  If an
old compact core is already present and later accreted gas dissipates and
settles near a finite radius, what is the minimal radial geometry needed
to maintain a depleted interval between the core and the outer ring?  A
related mechanical problem was studied by Bannikova, who considered a
central mass plus a massive ring or torus and showed that the
competition of their gravitational fields can create an OSCO--LC region
where stable circular motion is absent \citep{Bannikova2018}.  Our
mechanism is complementary but different.  We do not require the ring to
be a fully self-gravitating torus.  Instead we ask whether the reduced
radial imprint of a settled shell can create a barrier--well structure
outside a compact inner boundary.

The answer is yes, but the topology is not the usual interior double
well.  In a regular effective potential, two detached zones would be
represented by
\begin{equation}
{\rm minimum}\;-\;{\rm maximum}\;-\;{\rm minimum},
\end{equation}
with both the core and the ring generated by internal critical points.
For a compact Kepler-like background the inner state is instead supplied
by the boundary.  The required sequence becomes
\begin{equation}
{\rm boundary}\;-\;{\rm maximum}\;-\;{\rm minimum}.
\end{equation}
This distinction is the central result of the {paper}: Hoag-like radial
layering can be organized by a boundary-supported shell-localized
bifurcation rather than by a fully interior double well.

\section{Shell-deformed Kepler control model.}
\subsection{{Stationary formulation.}}
Let
\begin{equation}
x=\frac{r}{R_{\rm s}}
\end{equation}
be a dimensionless radius, where $r$ is the physical galactocentric
radius and $R_{\rm s}$ is the radial scale used to nondimensionalize the
model.  In the observational comparison below, $R_{\rm s}$ is fixed by
matching the model ring-supporting radius to the observed ring radius,
$R_{\rm s}=R_{\rm ring}^{\rm obs}/x_{\rm r}$.  The control potential is
\begin{equation}
\Phi(x)=
-\frac{\mu}{x}
+\epsilon
\exp\!\left[-\frac{(x-x_0)^2}{2\Delta^2}\right],
\qquad \mu>0 .
\label{eq:phi}
\end{equation}
Here $x_0$, $\Delta$, and $\epsilon$ describe the
settling radius, radial width, and control strength of the shell.
The Kepler term represents the compact-core control limit.  The Gaussian
term is an effective reduced deformation, not a Poisson-self-consistent
gas potential; it represents the radial imprint of externally supplied
material that has dissipated, partially circularized, and accumulated
near a finite radius. 

We diagnose radial barriers and wells using
\begin{equation}
W(x;\ell)=e^{2\Phi(x)}
\left(1+\frac{\ell^2}{x^2}\right),
\label{eq:W}
\end{equation}
where $\ell$ labels the angular-momentum scale of material participating
in the shell-supported structure.  This form can be motivated by an
auxiliary radial geodesic construction.  For the effective line element
\begin{equation}
\dd s^2=-e^{2\Phi(x)}\dd t^2+e^{-2\Phi(x)}\dd x^2+x^2\dd\varphi^2 ,
\end{equation}
timelike curves have conserved labels
$E=e^{2\Phi}\dot t$ and $\ell=x^2\dot\varphi$, and the normalization
condition gives
\begin{equation}
\dot x^2+
e^{2\Phi(x)}
\left(1+\frac{\ell^2}{x^2}\right)
=E^2 .
\end{equation}
Equation (\ref{eq:W}) is therefore a reduced square-energy function.  It
is used only as an effective radial control diagnostic; in the
weak-field limit
\begin{equation}
W=1+2\Phi+\frac{\ell^2}{x^2}
+O\!\left(\Phi^2,\frac{\Phi\ell^2}{x^2}\right),
\end{equation}
so its extrema track the leading Newtonian barrier--well ordering.

Critical radii satisfy $W'(x_\ast)=0$, equivalently
\begin{equation}
\frac{\dd\Phi}{\dd x}=
\frac{\ell^2}{x(x^2+\ell^2)} .
\label{eq:critical}
\end{equation}
A local maximum of $W$ is interpreted as a radial barrier and a local
minimum as a supporting well.  The Kepler boundary gives
\begin{equation}
W(x)\rightarrow0\quad (x\rightarrow0),
\qquad
W(x)\rightarrow1\quad (x\rightarrow\infty).
\label{eq:asymptotic}
\end{equation}
Thus the central core state is not an interior critical point of
$W(x)$; it is represented by the compact inner boundary.  The shell
needs to create only the gap barrier and the outer ring well.

\subsection{{Model scope and time-dependent embedding.}}

The Gaussian term in Eq.~(\ref{eq:phi}) is a phenomenological control
deformation.  It is not derived from a density distribution through
Poisson's equation, and its amplitude $\epsilon$ should not be identified
directly with a shell mass.  Likewise, extrema of $W$ identify barrier-like and
well-like regions of the reduced radial landscape; they neither specify a
stellar distribution function nor establish nonlinear dynamical stability.
The claims of the present model are therefore restricted to the existence,
topology, robustness, and observable radial scaling of the boundary-supported
branch.  A self-consistent gas-dynamical or distribution-function realization
is a separate problem.

To distinguish the stationary calculation from the qualitative formation
picture, we introduce a general time-dependent embedding,
\begin{equation}
\Phi(x,\tau)=-\frac{\mu}{x}+\Phi_{\rm sh}(x,\tau),
\label{eq:phi_time_embedding}
\end{equation}
where the shell deformation evolves towards the stationary endpoint,
\begin{equation}
\begin{gathered}
\lim_{\tau\rightarrow\infty}\Phi_{\rm sh}(x,\tau)
=\epsilon\exp\!\left[-\frac{(x-x_0)^2}{2\Delta^2}\right],\\
\lim_{\tau\rightarrow\infty}\Phi(x,\tau)
=\Phi_{\rm f}(x)\equiv\Phi(x).
\end{gathered}
\label{eq:phi_stationary_limit}
\end{equation}
No particular growth law, nor separate evolution of the shell amplitude,
centre, or width, is assumed.  All critical-point, saddle-node,
regularization, and Monte Carlo results in this paper refer to the stationary
endpoint $\Phi_{\rm f}(x)$.  The time-dependent form only embeds that endpoint
in a minimal dissipative settling picture.

Gas parcels carrying the angular-momentum label $\ell_i$ may then be
represented, in the weak-field limit, by

\begin{equation}
U_{\rm eff}(x;\ell_i,\tau)
=
\Phi(x,\tau)+\frac{\ell_i^2}{2x^2}.
\label{eq:Ueff_embedding}
\end{equation}

and

\begin{equation}
\frac{\dd^2x_i}{\dd\tau^2}
+
\Gamma\frac{\dd x_i}{\dd\tau}
=
-\partial_x U_{\rm eff}(x_i;\ell_i,\tau),
\label{eq:damped_future}
\end{equation}

where $\Gamma>0$ represents radial-energy loss.  The corresponding radial
energy,

\begin{equation*}
E_{r,i}=\frac{1}{2}
\left(\frac{\dd x_i}{\dd\tau}\right)^2
+U_{\rm eff}(x_i;\ell_i,\tau),
\end{equation*}

obeys

\begin{equation}
\frac{\dd E_{r,i}}{\dd\tau}
=
-\Gamma
\left(\frac{\dd x_i}{\dd\tau}\right)^2
+
\partial_\tau U_{\rm eff}.
\label{eq:energy_embedding}
\end{equation}

When the control deformation evolves sufficiently slowly for a quasistatic
description to apply, gas in the outer basin can lose radial energy and
approach the ring-supporting minimum, whereas the intervening maximum is a
low-residence barrier.  Equations~(\ref{eq:phi_time_embedding})--
(\ref{eq:energy_embedding}) are introduced solely to provide a qualitative
embedding of the stationary endpoint in a minimal dissipative settling
scenario.  They are not solved, calibrated, or used in any of the quantitative
results reported in this paper.

\section{Boundary-supported saddle-node threshold.}
Writing Eq.~(\ref{eq:critical}) as $F(x)=0$ gives
\begin{equation}
F(x)=A(x)-\epsilon G(x),
\label{eq:F}
\end{equation}
where
\begin{equation}
A(x)=\frac{\mu}{x^2}
-\frac{\ell^2}{x(x^2+\ell^2)}
\end{equation}
and
\begin{equation}
G(x)=
\frac{x-x_0}{\Delta^2}
\exp\!\left[-\frac{(x-x_0)^2}{2\Delta^2}\right].
\end{equation}
The birth of a shell-induced maximum--minimum pair is a saddle-node
threshold of this one-dimensional critical-point equation
\citep{Strogatz2015,Arnold1986}.  At threshold,
\begin{equation}
F(x_{\rm c})=0,\qquad F'(x_{\rm c})=0,
\end{equation}
or, equivalently,
\begin{equation}
A'(x_{\rm c})G(x_{\rm c})-A(x_{\rm c})G'(x_{\rm c})=0,
\qquad
\epsilon_{\rm c}=\frac{A(x_{\rm c})}{G(x_{\rm c})}.
\label{eq:threshold}
\end{equation}

Here
\begin{equation}
A'(x)=
-\frac{2\mu}{x^3}
+
\frac{\ell^2(3x^2+\ell^2)}
{x^2(x^2+\ell^2)^2},
\label{eq:supp_Aprime}
\end{equation}
and
\begin{equation}
G'(x)=
\frac{1}{\Delta^2}
\exp\!\left[-\frac{(x-x_0)^2}{2\Delta^2}\right]
\left[
1-\frac{(x-x_0)^2}{\Delta^2}
\right].
\label{eq:supp_Gprime}
\end{equation}
Equation~(\ref{eq:threshold}) is the semi-analytic threshold
surface that organizes the Monte Carlo branch in the {paper}.

At a critical point, the sign of $W''$ determines whether the point is a
maximum or a minimum.  Equivalently, because
\begin{equation}
W'(x)=2W(x)F(x),
\end{equation}
one has
\begin{equation}
W''(x_\ast)=2W(x_\ast)F'(x_\ast)
\end{equation}
at a root $F(x_\ast)=0$.  Thus $F'(x_\ast)<0$ gives a local maximum and
$F'(x_\ast)>0$ gives a local minimum.

The Monte Carlo branch below is therefore organized by an analytic
threshold surface rather than by an accidental numerical root pattern.

After the pair forms, let
\begin{equation}
x_{\rm g}<x_{\rm r}
\end{equation}
denote the maximum and subsequent minimum:
\begin{equation}
\begin{gathered}
W'(x_{\rm g})=0,\quad W''(x_{\rm g})<0,\\
W'(x_{\rm r})=0,\quad W''(x_{\rm r})>0 .
\end{gathered}
\label{eq:maxmin}
\end{equation}
The maximum $x_{\rm g}$ is the gap barrier and the minimum $x_{\rm r}$
is the ring-supporting well.  On a finite grid we define
$W_{\rm in}=W(x_{\min})$, $W_{\rm out}=W(x_{\max})$,
$W_{\rm g}=W(x_{\rm g})$, and $W_{\rm r}=W(x_{\rm r})$.  A candidate is
boundary-supported if
\begin{equation}
W_{\rm in}<W_{\rm r}<W_{\rm g},
\qquad
W_{\rm r}<W_{\rm out}.
\label{eq:boundary}
\end{equation}
To ensure that the pair is genuinely driven by the shell, we impose
\begin{equation}
0.8\le\frac{x_{\rm g}}{x_0}\le1.2,
\qquad
1.0\le\frac{x_{\rm r}}{x_0}\le2.5 .
\label{eq:localization}
\end{equation}

The singular Kepler form should be understood as the compact-core limit.
A regularized core,
\begin{equation}
\Phi_{\rm c,reg}(x)=
-\frac{\mu}{(x^2+x_{\rm c}^2)^{1/2}},
\end{equation}
differs from $-\mu/x$ by $O(x_{\rm c}^2/x^3)$ outside the unresolved
central region.  Since accepted shells satisfy $x_{\rm g}\sim x_0$ with
$x_0=O(1)$, the shell-induced roots are perturbatively unchanged for
$x_{\rm c}\ll x_{\rm g}$.  The boundary condition is then imposed at an
inner numerical radius $x_{\rm in}\sim x_{\rm c}$ by requiring
$W(x_{\rm in})<W_{\rm r}$.  This formulation makes clear that the
mechanism depends on a compact inner boundary, not on treating the
mathematical singularity as a literal galactic core.

\subsection{{Compact-Core Regularization}}

The numerical retention test used the regularized potential above

and imposed the inner boundary at
\begin{equation}
x_{\rm in}=\max(10^{-3},x_c).
\label{eq:supp_xin_reg}
\end{equation}
The critical-point function becomes
\begin{equation}
F_{\rm reg}(x)=
\frac{\mu x}{(x^2+x_c^2)^{3/2}}
-\epsilon G(x)
-\frac{\ell^2}{x(x^2+\ell^2)}.
\label{eq:supp_F_reg}
\end{equation}
The derivative used for classification is
\begin{equation}
F_{\rm reg}'(x)=
\frac{\mu(x_c^2-2x^2)}{(x^2+x_c^2)^{5/2}}
-\epsilon G'(x)
+
\frac{\ell^2(3x^2+\ell^2)}
{x^2(x^2+\ell^2)^2}.
\label{eq:supp_Freg_prime}
\end{equation}

We performed a retention test on the original $N_{\rm loc}=5494$
localized candidates.  For each $x_c$, we reclassified the roots using
Eq.~(\ref{eq:supp_F_reg}) and required the same shell-localized
boundary-supported conditions, now with $W(x_{\rm in})<W_{\rm r}$.  The
results are shown in Table~\ref{tab:supp_reg_core}.  The accepted branch
is unchanged for $x_c\le0.05$ and retains $99.5\%$ of candidates even at
$x_c=0.10$, while the median and central interval of $q=x_{\rm g}/x_{\rm r}$
remain unchanged at the quoted precision.

\begin{table}
\caption{Retention of the original localized branch under compact-core
regularization.}
\label{tab:supp_reg_core}
\begin{tabular}{ccccc}
$x_c$ & $x_{\rm in}$ & $N_{\rm retained}$ & $f_{\rm retained}$ &
$q_{16}$--$q_{50}$--$q_{84}$ \\
\hline
0 & $10^{-3}$ & 5494 & 1.0000 & 0.677--0.818--0.906 \\
0.01 & 0.01 & 5494 & 1.0000 & 0.677--0.818--0.906 \\
0.03 & 0.03 & 5494 & 1.0000 & 0.677--0.818--0.906 \\
0.05 & 0.05 & 5494 & 1.0000 & 0.677--0.818--0.906 \\
0.10 & 0.10 & 5468 & 0.9953 & 0.677--0.818--0.906 \\
\end{tabular}
\end{table}

\begin{figure}
\centering
\includegraphics[width=\linewidth]{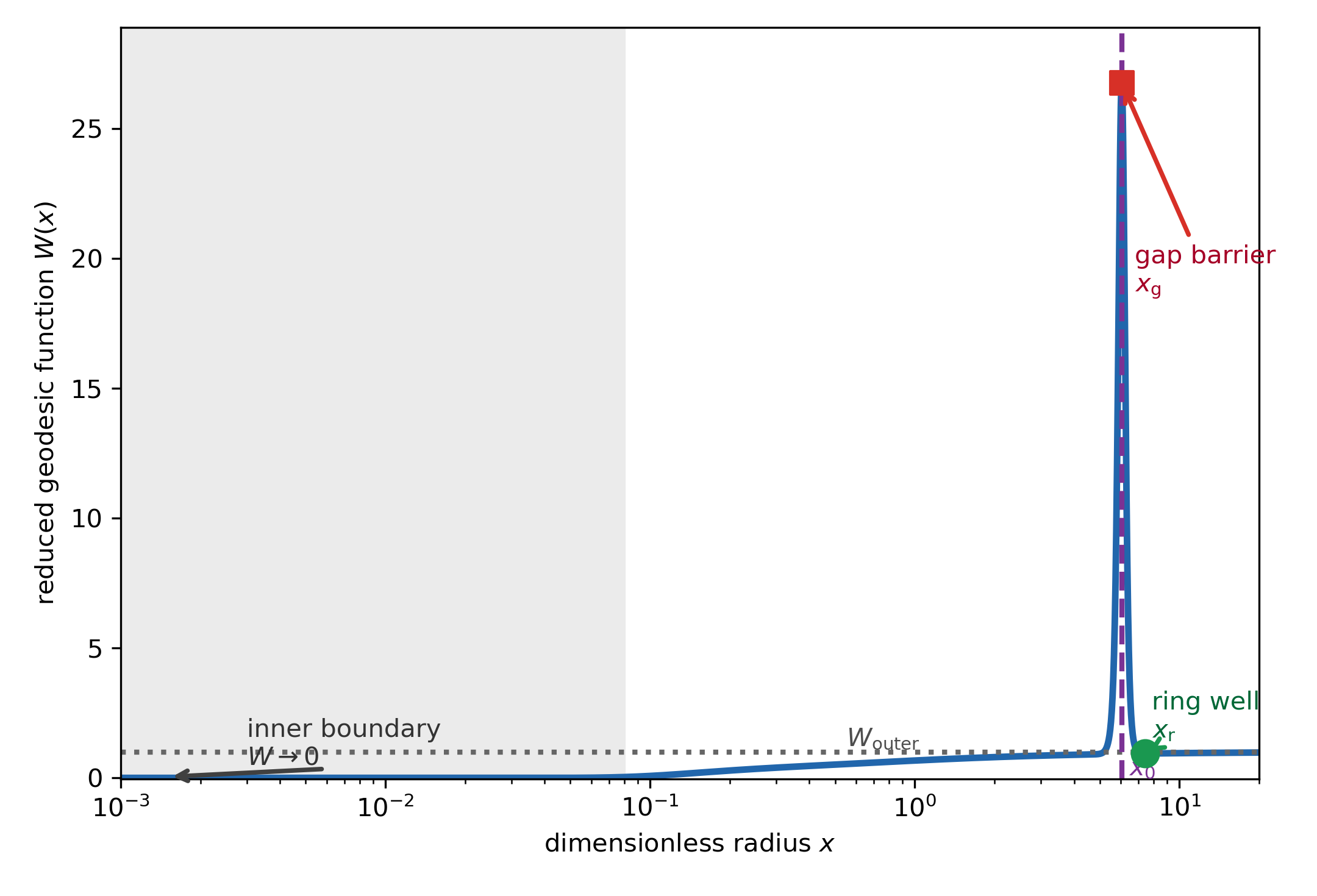}
\caption{
Boundary-supported radial-layering mechanism.  The compact inner
boundary supplies the core state.  A localized shell centered at $x_0$
creates an internal maximum $x_{\rm g}$ and an outer minimum $x_{\rm r}$,
identified with the gap barrier and ring-supporting well.
}
\par\smallskip
\noindent\textbf{Alt text:} A radial potential curve remains near zero
at small dimensionless radius, rises sharply at the shell location
$x_0$, reaches a narrow maximum labelled as the gap barrier
$x_{\rm g}$, and then falls to an outer minimum labelled as the
ring-supporting well $x_{\rm r}$.
\label{fig:mechanism}
\end{figure}

\section{Monte Carlo evidence.}
We scanned the five-dimensional parameter space
\begin{equation}
(\mu,\ell,x_0,\Delta,\epsilon)
\end{equation}
using $10^4$ random Monte Carlo realizations.  The ranges
were
\begin{equation}
\begin{gathered}
\mu\in[0.1,3],\quad
\ell\in[0.3,3],\quad
x_0\in[2,8],\\
\Delta\in[0.1,1],\quad
\epsilon\in[0,2],
\end{gathered}
\end{equation}
with log-uniform sampling for $\mu$, $\ell$, and $\Delta$, uniform
sampling for $x_0$ and $\epsilon$, and $\Delta/x_0\le0.5$.
For each realization all internal critical points of $W$ were located
and classified.   The full numerical recipe, grid convergence
checks, object-level filtering, and the Monte Carlo catalogue format are
incorporated below and are also provided in the public reproducibility
repository.

Interior minimum--maximum--minimum subsequences can occur in the full
sample, usually as part of profiles with additional inner structure.
However, they are disjoint from the shell-localized Hoag-like branch:
none of the localized candidates contains an ordered interior
minimum--maximum--minimum subsequence.  By contrast, internal
maximum--minimum pairs occurred with
fraction
\begin{equation}
f_{\rm max-min}=0.5551 ,
\end{equation}
and every such realization satisfied the boundary inequalities in
Eq.~(\ref{eq:boundary}) within the sampled domain.  After applying the
localization condition in Eq.~(\ref{eq:localization}), we obtained
\begin{equation}
N_{\rm loc}=5494,\qquad f_{\rm loc}=0.5494 .
\end{equation}
Thus the dominant Hoag-like branch is not an ordinary interior double
well hidden in a small corner of parameter space.  It is a
shell-localized maximum--minimum pair supported by the compact inner
boundary.

\begin{figure}
\centering
\includegraphics[width=\linewidth]{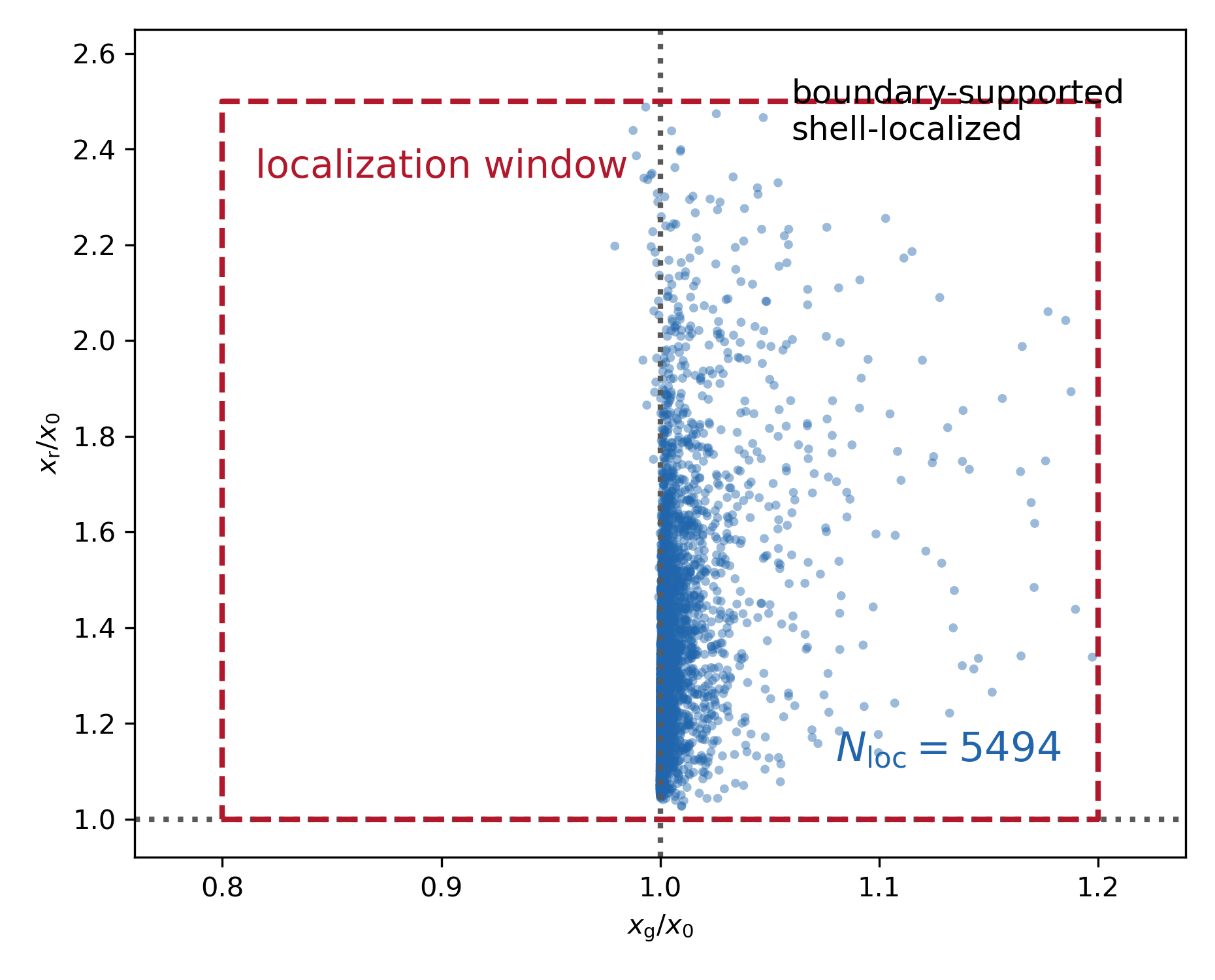}
\caption{
Monte Carlo evidence for shell localization.  Boundary-supported
candidates cluster near $x_{\rm g}/x_0\simeq1$, while the ring-supporting
minimum forms outside the shell.  Sparse nonlocal roots are removed by
Eq.~(\ref{eq:localization}).
}
\par\smallskip
\noindent\textbf{Alt text:} Scatter plot of Monte Carlo candidates in
the $x_{\rm g}/x_0$--$x_{\rm r}/x_0$ plane.  Most blue points lie inside
the localization window, concentrated just to the right of
$x_{\rm g}/x_0=1$, with $x_{\rm r}/x_0$ greater than one.
\label{fig:localization}
\end{figure}

Figure~\ref{fig:localization} shows the corresponding localization
diagnostic.  Accepted candidates cluster near
$x_{\rm g}/x_0\simeq1$, demonstrating that the gap barrier tracks the
settled shell.  The ring well lies outside the shell, typically with
$x_{\rm r}/x_0>1$.  This behavior is the key difference from a generic
curve-fitting potential: the model identifies where the gap should be
relative to the shell.

\subsection{{Root Classification and Monte Carlo Scan}}

The numerical classification used a logarithmic radial grid,
\begin{equation}
x\in[10^{-3},30],
\qquad
N_x=5000.
\label{eq:supp_grid}
\end{equation}
Roots of $F(x)$ were identified from sign changes on the grid and
refined by one-dimensional bracketing.  Critical points were classified
by the sign of $F'$ or, equivalently, $W''$.  This procedure classifies
the reduced radial profile; it is not a distribution-function
calculation.

The Monte Carlo scan used
\begin{equation}
\bm{\theta}=(\mu,\ell,x_0,\Delta,\epsilon).
\end{equation}
The ranges and priors are listed in Table~\ref{tab:supp_priors}.  The
random seed was 1234 and the total number of realizations was
$N_{\rm tot}=10^4$.  We imposed $\Delta/x_0\le0.5$ so that the shell
remained localized relative to its center.

\begin{table}
\caption{Monte Carlo parameter ranges.}
\label{tab:supp_priors}
\begin{tabular}{lccc}
Parameter & Range & Prior & Meaning \\
\hline
$\mu$ & $[0.1,3]$ & log-uniform & compact-core strength \\
$\ell$ & $[0.3,3]$ & log-uniform & angular-momentum label \\
$x_0$ & $[2,8]$ & uniform & shell center \\
$\Delta$ & $[0.1,1]$ & log-uniform & shell width \\
$\epsilon$ & $[0,2]$ & uniform & shell amplitude \\
\end{tabular}
\end{table}

For each realization we recorded whether the profile contained an exact
three-critical-point interior min--max--min sequence, any ordered
interior min--max--min subsequence, an internal max--min pair, a
boundary-supported candidate, and a shell-localized boundary-supported
candidate.  This distinction matters because min--max--min subsequences
can occur in the full scan as part of profiles with additional inner
structure, but they are disjoint from the shell-localized Hoag-like
branch.  The shell-localization window was
\begin{equation}
0.8\le \frac{x_{\rm g}}{x_0}\le1.2,
\qquad
1.0\le \frac{x_{\rm r}}{x_0}\le2.5.
\label{eq:supp_local_window}
\end{equation}
The occupancy results are summarized in Table~\ref{tab:supp_occupancy}.

\begin{table}
\caption{Occupancy of structural classes in the $10^4$-point scan.}
\label{tab:supp_occupancy}
\resizebox{\columnwidth}{!}{%
\begin{tabular}{lcc}
Class & Number & Fraction \\
\hline
Exact three-critical-point min--max--min & 0 & 0 \\
Ordered min--max--min subsequence in full sample & 3988 & 0.3988 \\
Internal max--min pair & 5551 & 0.5551 \\
Boundary-supported candidate & 5551 & 0.5551 \\
Shell-localized boundary-supported & 5494 & 0.5494 \\
Ordered min--max--min subsequence within localized branch & 0 & 0 \\
\end{tabular}
}
\end{table}

The equality between the internal max--min and boundary-supported
fractions is a consequence of the Kepler asymptotic structure in the
sampled domain.  Once a shell-induced max--min pair is present, the
inner boundary and outer plateau supply the required bounding levels.

\begin{figure}
\centering
\includegraphics[width=\linewidth]{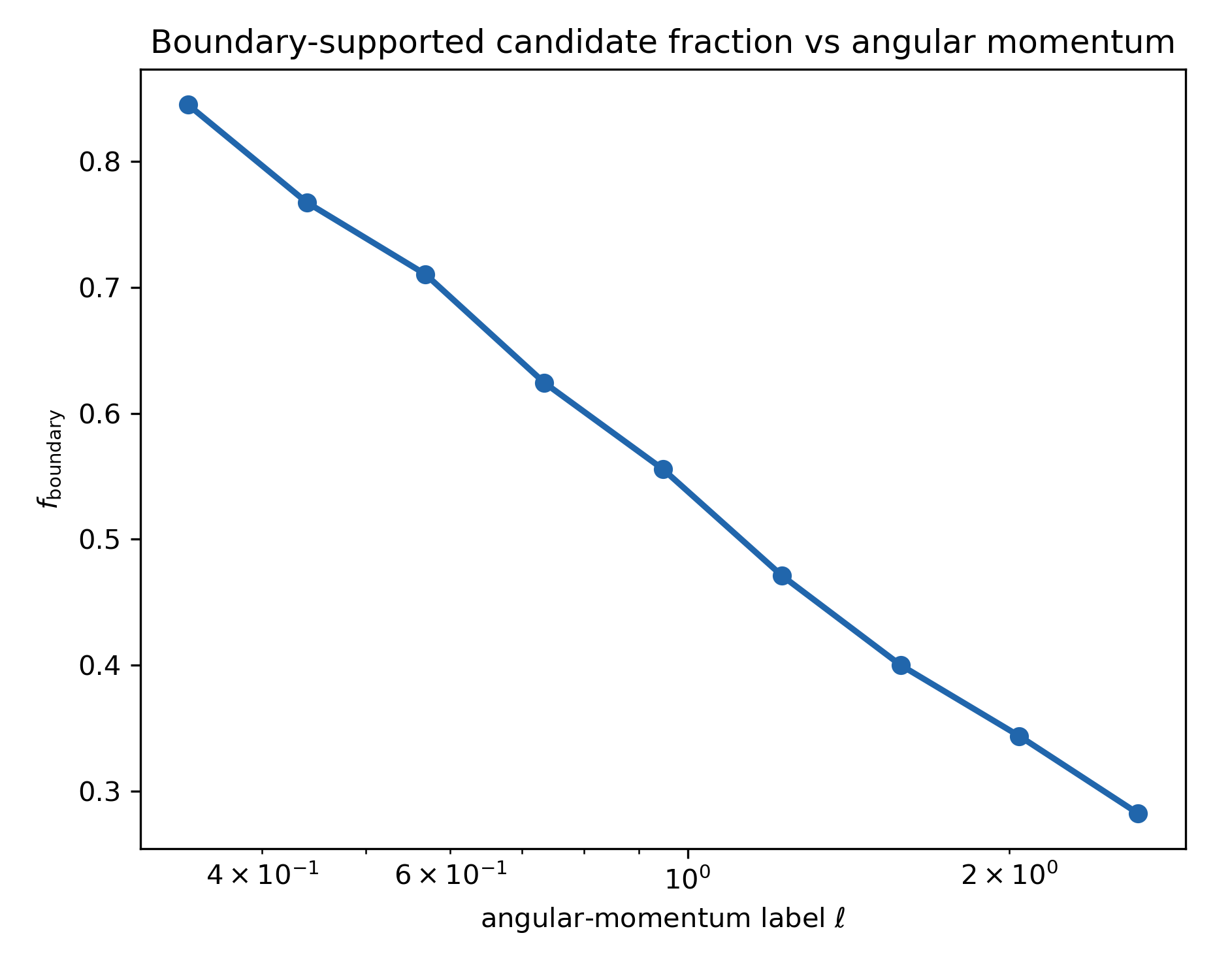}
\caption{
Boundary-supported candidate fraction as a function of angular-momentum
label $\ell$.  The decline with increasing $\ell$ shows that
low-to-moderate angular momentum is favored for the clean
shell-localized branch.
}
\par\smallskip
\noindent{\textbf{Alt text:} A line plot shows the
boundary-supported candidate fraction decreasing steadily from about
$0.85$ at low $\ell$ to about $0.28$ at the largest sampled $\ell$.}
\label{fig:supp_fboundary_ell}
\end{figure}

\subsection{{Diagnostics of the Shell-Localized Branch}}

For each shell-localized candidate we stored the dimensionless radii
$x_{\rm g}$, $x_{\rm r}$, and $x_0$, the ratios $x_{\rm g}/x_0$ and
$x_{\rm r}/x_0$, and the scale-free morphology ratio
\begin{equation}
q_{\rm model}
\equiv
\frac{x_{\rm g}}{x_{\rm r}}
=
\frac{R_{\rm gap}}{R_{\rm ring}}.
\label{eq:supp_qmodel}
\end{equation}
We also recorded the contrast diagnostics
\begin{equation}
\begin{aligned}
\Delta W_{\rm g-r}=W_{\rm g}-W_{\rm r},
\qquad &
\Delta W_{\rm out-r}=W_{\rm out}-W_{\rm r},
\\
&
\Delta W_{\rm r-in}=W_{\rm r}-W_{\rm in}.
\end{aligned}
\end{equation}
Summary statistics are listed in Table~\ref{tab:supp_localized_stats}.
The accepted branch has $x_{\rm g}/x_0\simeq1$, while the ring well lies
moderately outside the shell.

\begin{table*}
\caption{Structural diagnostics of the shell-localized branch.}
\label{tab:supp_localized_stats}
\begin{tabular}{lccc}
Quantity & 16\% & Median & 84\% \\
\hline
$\mu$ & 0.474 & 1.154 & 2.214 \\
$\ell$ & 0.382 & 0.681 & 1.572 \\
$x_0$ & 2.946 & 5.004 & 6.959 \\
$\Delta$ & 0.146 & 0.313 & 0.698 \\
$\epsilon$ & 0.360 & 1.025 & 1.683 \\
$x_{\rm g}$ & 2.973 & 5.022 & 6.974 \\
$x_{\rm r}$ & 4.065 & 6.251 & 8.364 \\
$x_{\rm g}/x_0$ & 1.000 & 1.001 & 1.008 \\
$x_{\rm r}/x_0$ & 1.106 & 1.227 & 1.489 \\
$x_{\rm g}/x_{\rm r}$ & 0.677 & 0.818 & 0.906 \\
$x_{\rm r}/x_{\rm g}$ & 1.104 & 1.222 & 1.477 \\
$\Delta W_{\rm g-r}$ & 0.538 & 3.831 & 16.786 \\
$\Delta W_{\rm out-r}$ & 0.094 & 0.209 & 0.372 \\
$\Delta W_{\rm r-in}$ & 0.502 & 0.715 & 0.872 \\
\end{tabular}
\end{table*}

\begin{figure}
\centering
\includegraphics[width=\linewidth]{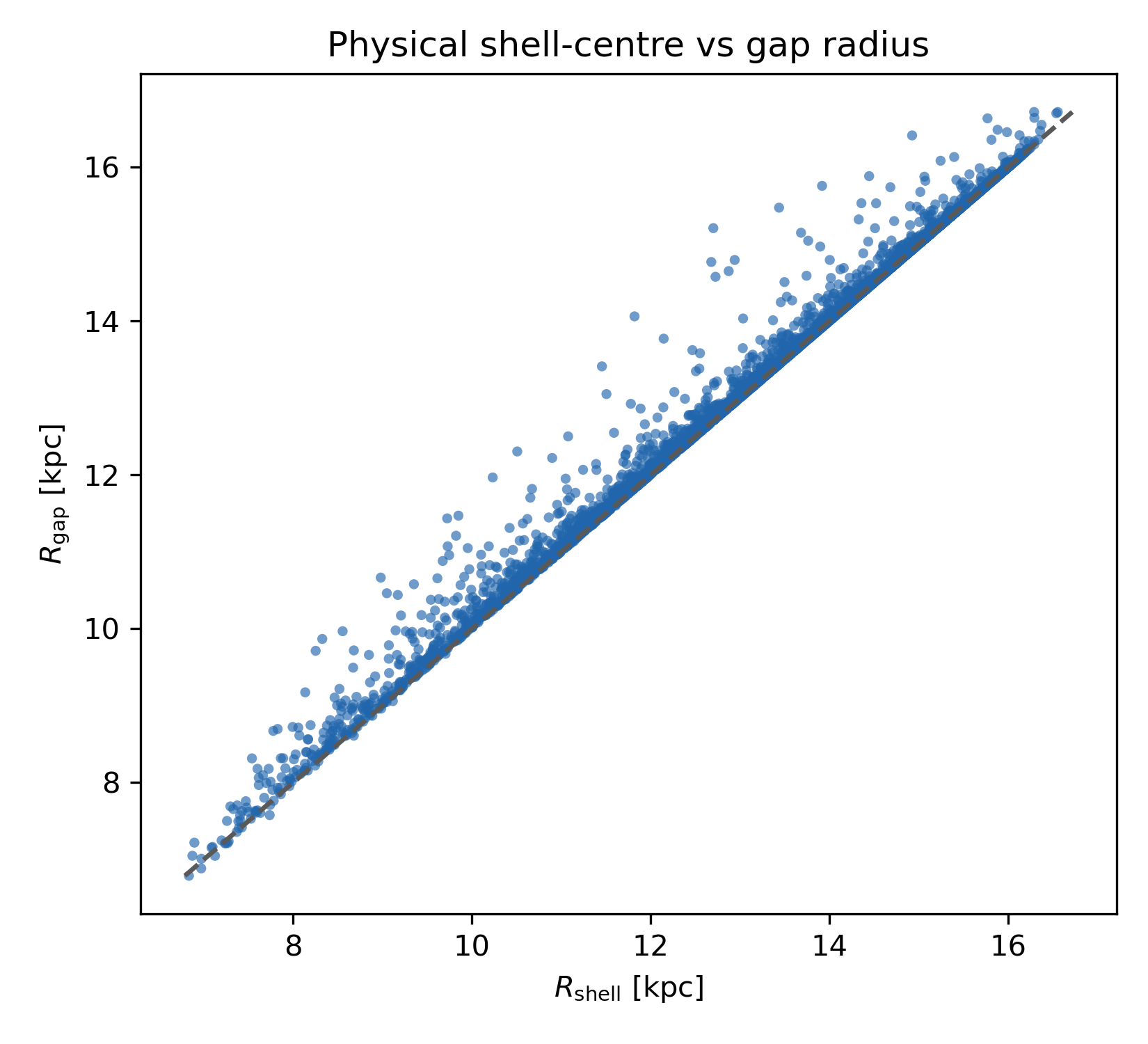}
\caption{
Physical shell-center radius versus physical gap radius after calibrating
the ring radius to $17\,{\rm kpc}$.  The near-diagonal relation shows
that the gap barrier is shell-tracking.
}
\par\smallskip
\noindent{\textbf{Alt text:} A scatter plot of physical shell-center
radius against physical gap radius forms a narrow near-diagonal sequence from
approximately $7$ to $17\,{\rm kpc}$ in both coordinates.}
\label{fig:supp_Rshell_Rgap}
\end{figure}

After fixing a physical ring radius,
\begin{equation}
\begin{aligned}
R_s=\frac{R_{\rm ring}^{\rm obs}}{x_{\rm r}},
\qquad &
R_{\rm gap}=R_sx_{\rm g},
\\
&
R_{\rm shell}=R_sx_0,
\qquad
\Delta_{\rm phys}=R_s\Delta.
\end{aligned}
\label{eq:supp_rescale}
\end{equation}
For $R_{\rm ring}^{\rm obs}=17\,{\rm kpc}$ the physical summaries are
given in Table~\ref{tab:supp_physical_stats}.  The near-diagonal
relation between $R_{\rm shell}$ and $R_{\rm gap}$ is shown in
Fig.~\ref{fig:supp_Rshell_Rgap}.

\begin{table}
\caption{Physical scaling for $R_{\rm ring}^{\rm obs}=17\,{\rm kpc}$.}
\label{tab:supp_physical_stats}
\begin{tabular}{lccc}
Quantity & 16\% & Median & 84\% \\
\hline
$R_s$ [kpc] & 2.033 & 2.719 & 4.182 \\
$R_{\rm gap}$ [kpc] & 11.508 & 13.906 & 15.402 \\
$R_{\rm shell}$ [kpc] & 11.420 & 13.855 & 15.369 \\
$\Delta_{\rm phys}$ [kpc] & 0.458 & 0.943 & 1.748 \\
\end{tabular}
\end{table}

\section{Observable scaling.}
For each localized candidate, an observed ring radius fixes
\begin{equation}
R_{\rm s}=\frac{R_{\rm ring}^{\rm obs}}{x_{\rm r}} .
\end{equation}
The model then gives
\begin{equation}
\frac{R_{\rm gap}}{R_{\rm ring}}
=
\frac{x_{\rm g}}{x_{\rm r}} ,
\label{eq:ratio}
\end{equation}
which is independent of the absolute distance scale.  We define
$R_{\rm ring}$ as the outer-ring peak radius, or the representative
mid-ring radius when a peak is not tabulated, and $R_{\rm gap}$ as the
surface-brightness minimum between core and ring, or the inner onset of
the detached ring when that minimum is not separately reported.  Because
these literature radii are heterogeneous, the comparison below is a
morphology-scale consistency test rather than a population-level fit.

The localized branch gives
\begin{equation}
\begin{gathered}
{\rm median}\left(\frac{R_{\rm gap}}{R_{\rm ring}}\right)=0.818,\\
16{\rm --}84\%:\;0.677{\rm --}0.906 .
\end{gathered}
\label{eq:model_interval}
\end{equation}
The representative observed ratios for Hoag's Object, UGC 4599, and PGC
1000714 are approximately
\begin{equation}
0.890,\qquad 0.780,\qquad 0.717,
\end{equation}
respectively
\citep{Schweizer1987,Finkelman2011Hoag,FinkelmanBrosch2011UGC4599,
MutluPakdil2017PGC1000714}.  All three fall within the central model
interval.  JO171, whose ring has been interpreted as externally formed
and is now being reshaped by ram-pressure stripping in a cluster
environment, has
\begin{equation}
\left(R_{\rm gap}/R_{\rm ring}\right)_{\rm JO171}\simeq0.325
\end{equation}
using the ring extent and inner rise radius from the MUSE analysis
\citep{Moretti2018JO171}; it lies outside the clean shell-localized
interval.  This is not a decisive exclusion test for JO171's formation
history, but it is consistent with treating JO171 as an environmentally
processed comparison object rather than a member of the isolated
shell-supported branch.

The gap and ring definitions, adopted uncertainty bands, Hoag A/B gap
choices, and velocity-normalized $j_{\rm phys}$ distributions are
tabulated below and supplied in machine-readable form in the public
reproducibility repository.

With $R_{\rm ring}^{\rm obs}=17\,{\rm kpc}$ as a fiducial Hoag-like
calibration, the localized branch gives median physical scales
\begin{equation}
\begin{gathered}
R_{\rm gap}\simeq13.9\,{\rm kpc},\qquad
R_{\rm shell}\simeq13.9\,{\rm kpc},\\
\Delta_{\rm phys}\simeq0.94\,{\rm kpc}.
\end{gathered}
\end{equation}
The near equality of $R_{\rm gap}$ and $R_{\rm shell}$ is the physical
expression of $x_{\rm g}\simeq x_0$: the depleted gap should trace the
settled shell radius.

\begin{figure}
\centering
\includegraphics[width=\linewidth]{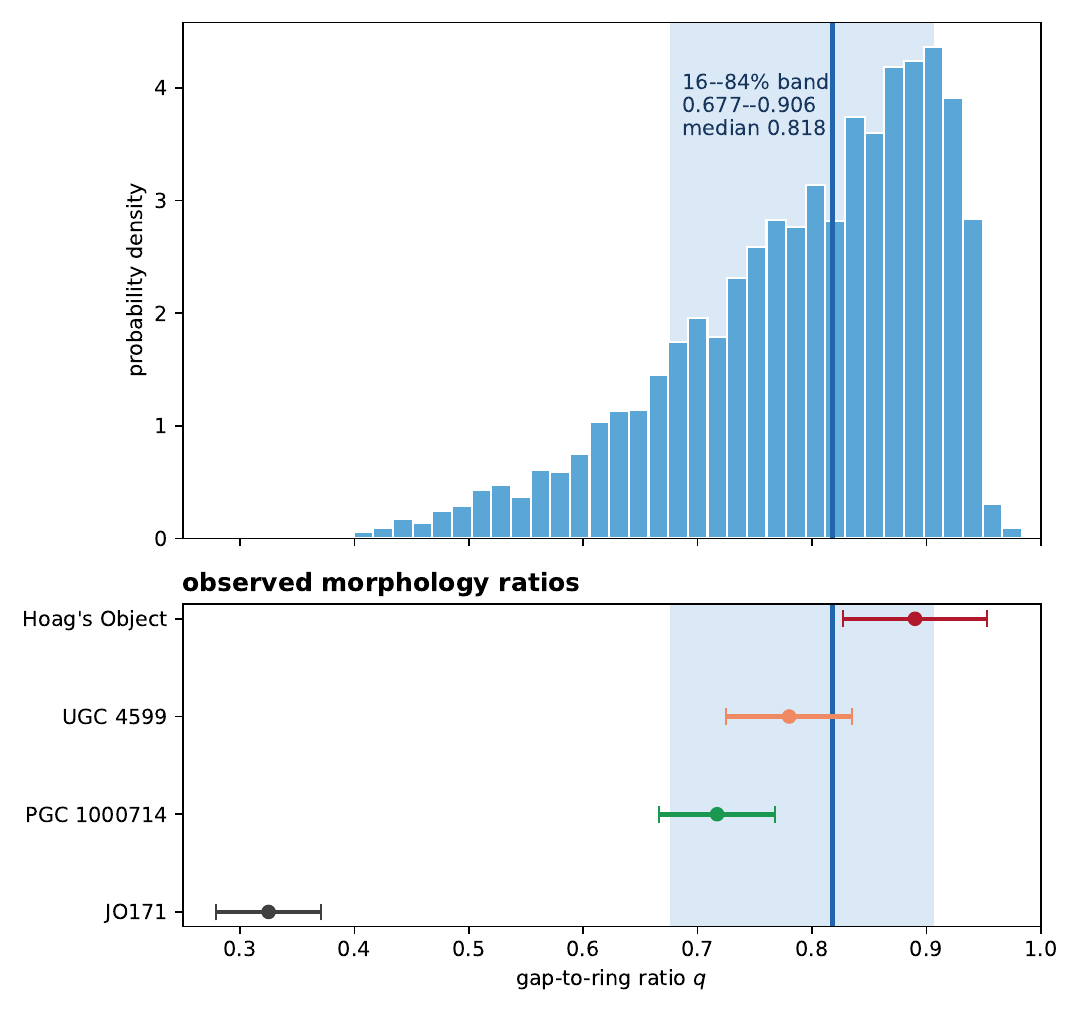}
\caption{
Model probability density and observed morphology ratios.  The upper
panel shows the probability density of
$q=R_{\rm gap}/R_{\rm ring}=x_{\rm g}/x_{\rm r}$ for the shell-localized
Monte Carlo branch.  The gray band marks the central $16$--$84\%$
interval and the dashed line marks the median.  The lower panel shows
representative observed ratios with conservative morphology-definition
uncertainties.  The three clean Hoag-like systems overlap the model
density interval, while JO171 lies outside it.
}
\par\smallskip
\noindent\textbf{Alt text:} Two-panel comparison of model and observed
gap-to-ring ratios.  The upper histogram peaks within the shaded
central interval from $q=0.677$ to $0.906$ with median $q=0.818$; the
lower panel shows Hoag, UGC~4599, and PGC~1000714 overlapping this
interval, while JO171 lies at much smaller $q$.
\label{fig:observed}
\end{figure}

\subsection{{Observed Ratios and Conservative Uncertainties}}

The observational comparison uses
\begin{equation}
q_{\rm obs}=
\frac{R_{\rm gap}^{\rm obs}}{R_{\rm ring}^{\rm obs}}.
\end{equation}
We define $R_{\rm ring}^{\rm obs}$ as the outer-ring peak radius, or a
representative mid-ring radius if no peak is tabulated.  We define
$R_{\rm gap}^{\rm obs}$ as the surface-brightness minimum between the
core and ring, or the inner onset of the detached ring when the minimum
is not separately reported.  Since the literature radii are not derived
from a homogeneous profile-fitting pipeline, we attach conservative
morphology-definition uncertainties rather than formal statistical
errors.

For
\begin{equation}
q=\frac{R_{\rm gap}}{R_{\rm ring}},
\end{equation}
we propagate radial uncertainties as
\begin{equation}
\sigma_q=
q
\left[
\left(\frac{\sigma_{\rm gap}}{R_{\rm gap}}\right)^2
+
\left(\frac{\sigma_{\rm ring}}{R_{\rm ring}}\right)^2
\right]^{1/2}.
\label{eq:supp_sigma_q}
\end{equation}
For Hoag's Object, UGC 4599, and PGC 1000714 we use $5\%$ uncertainties
on both adopted radii.  For JO171 we use $10\%$ because its quoted ring
scale is an outer extent rather than a homogeneous ring-peak radius.
The resulting values are listed in Table~\ref{tab:supp_obs_uncert}.

\begin{table*}
\caption{Observed morphology ratios and conservative definition
uncertainties.  The quoted uncertainties are not homogeneous measurement
errors; they quantify sensitivity to the adopted gap and ring radius
definitions.}
\label{tab:supp_obs_uncert}
\begin{tabular}{lcccccc}
Object & $R_{\rm gap}$ & $\sigma_{\rm gap}$ & $R_{\rm ring}$ &
$\sigma_{\rm ring}$ & $q_{\rm obs}$ & $\sigma_q$ \\
\hline
Hoag's Object & $17.0''$ & $0.85''$ & $19.1''$ & $0.96''$ & 0.890 & 0.063 \\
UGC 4599 & $39''$ & $1.95''$ & $50''$ & $2.50''$ & 0.780 & 0.055 \\
PGC 1000714 & $21.5''$ & $1.08''$ & $30''$ & $1.50''$ & 0.717 & 0.051 \\
JO171 & $6.5\,{\rm kpc}$ & $0.65\,{\rm kpc}$ &
$20\,{\rm kpc}$ & $2.0\,{\rm kpc}$ & 0.325 & 0.046 \\
\end{tabular}
\end{table*}

\subsection{{Object-Level Filtering}}

For each object we applied the filter
\begin{equation}
\left|q_{\rm model}-q_{\rm obs}\right|\le0.05,
\label{eq:supp_object_filter}
\end{equation}
where $q_{\rm model}=x_{\rm g}/x_{\rm r}$.  The purpose is not to infer
a unique best-fitting model, but to test whether each observed
morphology selects a finite region of the same localized branch.
The results are summarized in Table~\ref{tab:supp_object_filter}.  JO171
is not included because its ratio lies outside the clean shell-localized
interval.

\begin{table*}
\caption{Object-level filtering of the shell-localized branch.}
\label{tab:supp_object_filter}
\resizebox{\textwidth}{!}{%
\begin{tabular}{lcccccccc}
Object & $q_{\rm obs}$ & $N_{\rm obj}$ & $f_{\rm obj|loc}$ &
median $\mu$ & median $\ell$ & median $x_0$ & median $\Delta$ &
median $\epsilon$ \\
\hline
Hoag's Object & 0.890 & 2192 & 0.399 & 1.239 & 0.687 & 5.558 & 0.175 & 0.957 \\
UGC 4599 & 0.780 & 1515 & 0.276 & 1.150 & 0.679 & 5.042 & 0.408 & 1.002 \\
PGC 1000714 & 0.717 & 1167 & 0.212 & 1.122 & 0.696 & 4.867 & 0.582 & 1.065 \\
\end{tabular}
}
\end{table*}

With physical ring calibrations, the same filters give finite
kiloparsec-scale subsets.  For Hoag's Object, the median filtered values
are $R_{\rm gap}=15.13\,{\rm kpc}$,
$R_{\rm shell}=15.11\,{\rm kpc}$,
$\Delta_{\rm phys}=0.54\,{\rm kpc}$, and
$j_{\rm phys}=140.37\,{\rm kpc\,km\,s^{-1}}$.  For UGC 4599, using
$R_{\rm ring}=5\,{\rm kpc}$ and $V_s=130\,{\rm km\,s^{-1}}$, the median
values are $R_{\rm gap}=3.92\,{\rm kpc}$,
$R_{\rm shell}=3.90\,{\rm kpc}$,
$\Delta_{\rm phys}=0.33\,{\rm kpc}$, and
$j_{\rm phys}=77.75\,{\rm kpc\,km\,s^{-1}}$.

\subsection{{Hoag's Object A/B Gap Definitions}}

For Hoag's Object we considered two gap definitions.  Definition A uses
the inner onset of the bright ring segment,
\begin{equation}
q_{\rm A}=\frac{17.0''}{19.1''}=0.890.
\end{equation}
Definition B uses a more conservative core-minimum gap scale,
\begin{equation}
q_{\rm B}=\frac{13.1''}{19.1''}=0.686.
\end{equation}
Each localized candidate was rescaled to $R_{\rm ring}=17\,{\rm kpc}$
and $V_s=70\,{\rm km\,s^{-1}}$.  The hard filter was
\begin{equation}
\begin{aligned}
|q_{\rm model}-q_{\rm obs}|\le0.05,
\qquad &
0.3\le\Delta_{\rm phys}/{\rm kpc}\le2.0,
\\
&
50\le\frac{j_{\rm phys}}{{\rm kpc\,km\,s^{-1}}}\le350.
\end{aligned}
\label{eq:supp_hoag_filter}
\end{equation}
The selected subsets are summarized in Table~\ref{tab:supp_hoag_ab}.

\begin{table*}
\caption{Hoag's Object forward-filtered subsets.}
\label{tab:supp_hoag_ab}
\begin{tabular}{lcc}
Quantity & Definition A & Definition B \\
\hline
$q_{\rm obs}$ & 0.890 & 0.686 \\
$N_{\rm selected}$ & 1693 & 638 \\
$f_{\rm selected}$ & 0.308 & 0.116 \\
median $\mu$ & 1.117 & 0.926 \\
median $\ell$ & 0.616 & 0.645 \\
median $x_0$ & 5.613 & 4.900 \\
median $\Delta$ & 0.184 & 0.654 \\
median $\epsilon$ & 0.940 & 1.210 \\
median $R_{\rm gap}$ [kpc] & 15.12 & 11.86 \\
median $R_{\rm shell}$ [kpc] & 15.10 & 11.79 \\
median $\Delta_{\rm phys}$ [kpc] & 0.54 & 1.56 \\
median $j_{\rm phys}$ [$\mathrm{kpc\,km\,s^{-1}}$] & 125.55 & 117.62 \\
\end{tabular}
\end{table*}

\subsection{{Specific-Angular-Momentum Proxy}}

The dimensionless angular-momentum label is mapped to a physical proxy by
\begin{equation}
j_{\rm phys}=R_sV_s\ell.
\label{eq:supp_jphys}
\end{equation}
This is a velocity-normalized consistency check, not a dynamical
inversion.  The results scale linearly with the adopted velocity
normalization.  Table~\ref{tab:supp_j_scan} gives a velocity sweep for
the fiducial $R_{\rm ring}=17\,{\rm kpc}$ calibration.

\begin{table*}
\caption{Velocity sweep for $j_{\rm phys}=R_sV_s\ell$ with
$R_{\rm ring}=17\,{\rm kpc}$.}
\label{tab:supp_j_scan}
\begin{tabular}{ccccc}
$V_s$ [km s$^{-1}$] & 16\% & Median & 84\% & Range \\
\hline
70 & 70.96 & 140.25 & 336.99 & 32.56--1309.66 \\
130 & 131.78 & 260.47 & 625.83 & 60.46--2432.22 \\
150 & 152.06 & 300.54 & 722.11 & 69.76--2806.40 \\
200 & 202.74 & 400.72 & 962.82 & 93.02--3741.87 \\
250 & 253.43 & 500.90 & 1203.52 & 116.27--4677.34 \\
300 & 304.12 & 601.08 & 1444.23 & 139.52--5612.81 \\
\end{tabular}
\end{table*}

\section{{Illustrative Morphology Proxy}}

\begin{figure}
\centering
\includegraphics[width=\linewidth]{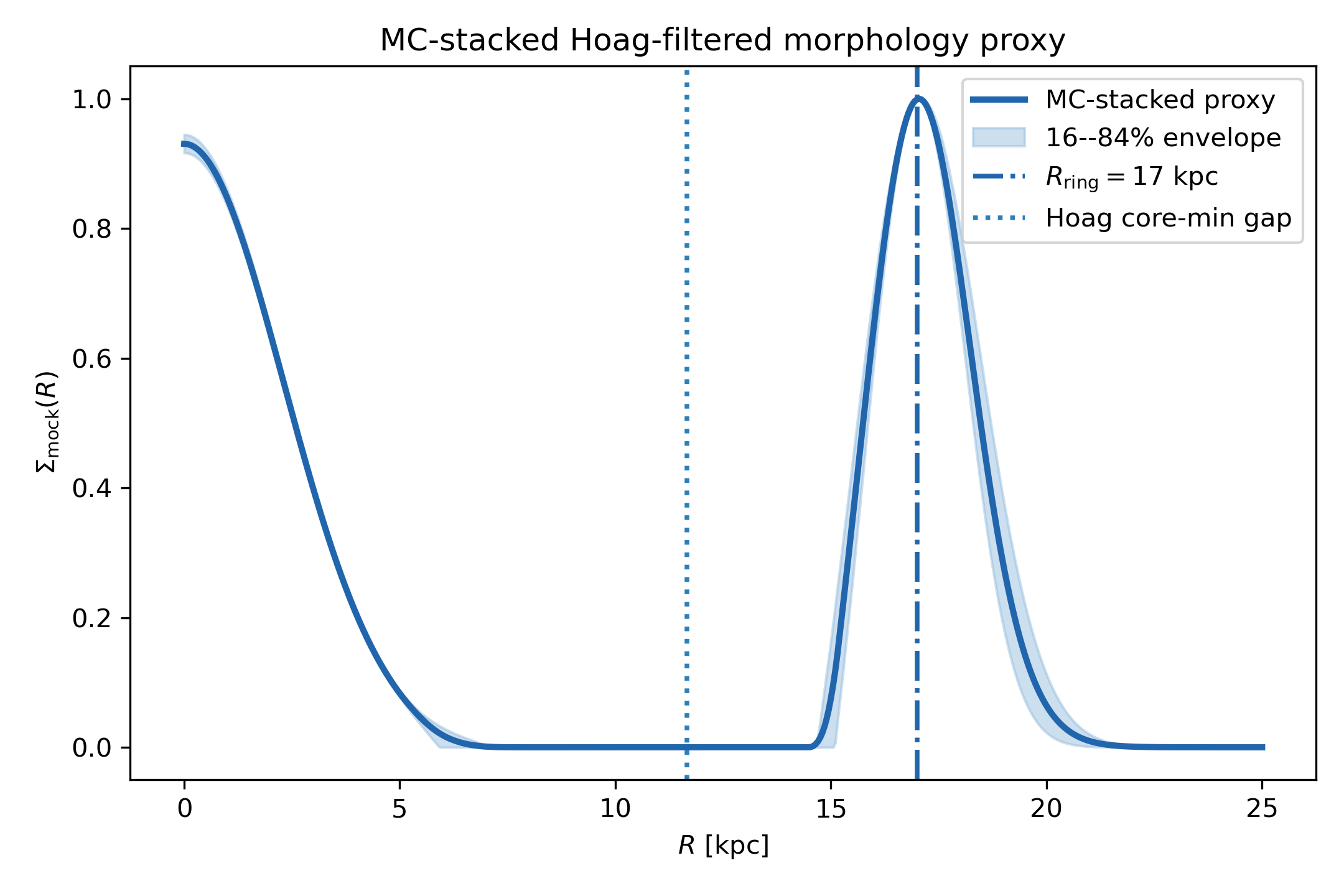}
\caption{
MC-stacked Hoag-filtered radial morphology proxy.  The stack shows a
compact central enhancement, an intermediate depleted region, and an
outer ring enhancement.
}
\par\smallskip
\noindent{\textbf{Alt text:} A radial profile shows a high compact
central component, a near-zero intermediate gap, and a narrow outer-ring peak
at $17\,{\rm kpc}$; a shaded band gives the central Monte Carlo envelope.}
\label{fig:supp_mock_profile}
\end{figure}

\begin{figure}
\centering
\includegraphics[width=\linewidth]{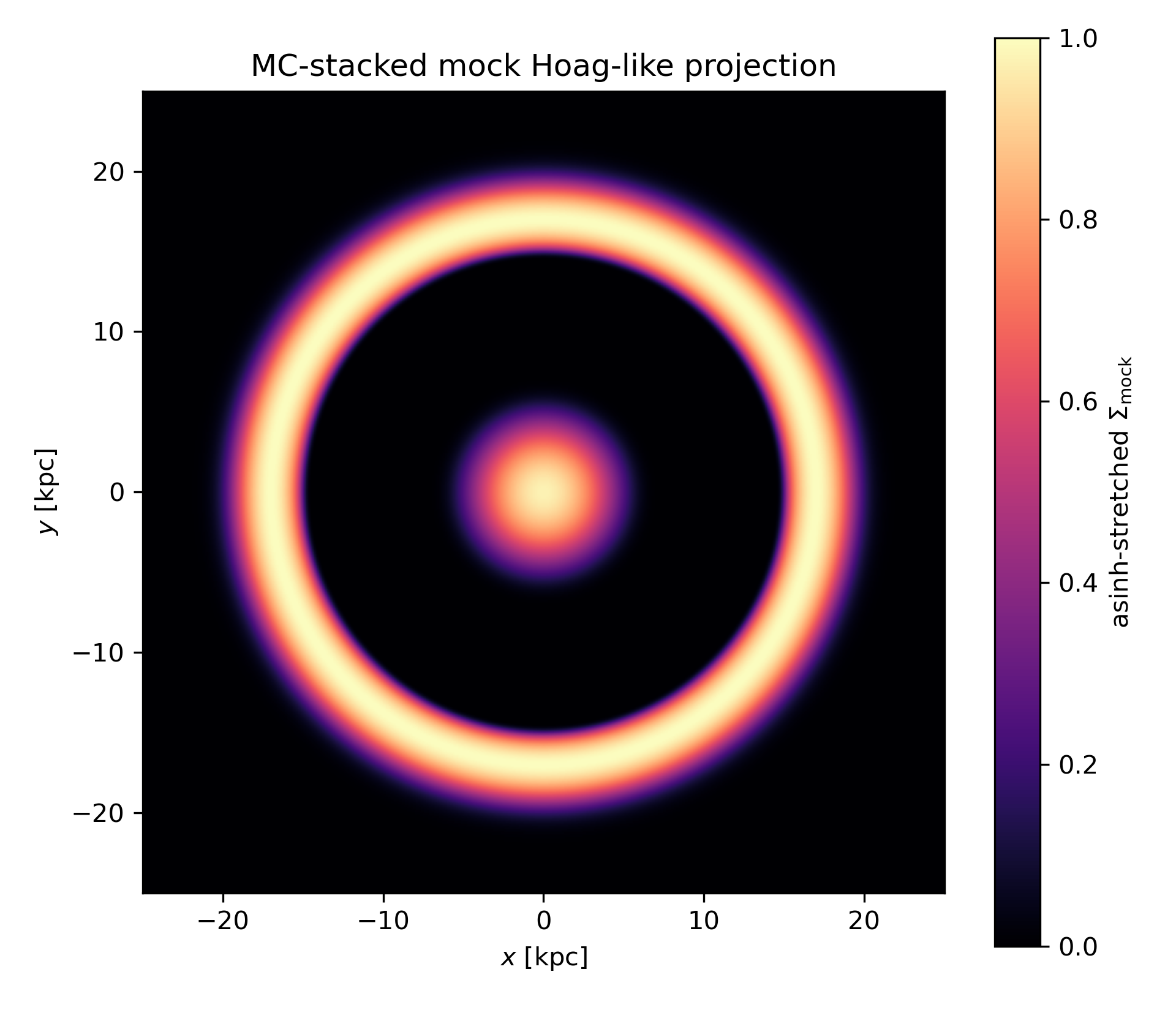}
\caption{
Axisymmetric MC-stacked mock projection corresponding to
Fig.~\ref{fig:supp_mock_profile}.  The image is intended only as a
qualitative visualization of the filtered branch.
}
\par\smallskip
\noindent{\textbf{Alt text:} A face-on intensity map shows a bright
compact central component, a dark annular gap, and a nearly circular luminous
outer ring centered at a radius of about $17\,{\rm kpc}$.}
\label{fig:supp_mock_projection}
\end{figure}

The mock projection is an illustrative morphology proxy only.  It is not
a distribution-function fit, a hydrodynamical calculation, or a
surface-brightness fit.  For a filtered candidate, we used the schematic
profile
\begin{equation}
\begin{aligned}
\Sigma_{\rm mock}(R)=
A_{\rm c}\exp\!\left[-\frac{R^2}{2\sigma_{\rm c}^2}\right]
&+
A_{\rm r}\exp\!\left[-\frac{(R-R_{\rm ring})^2}{2\sigma_{\rm r}^2}\right]
\\
&-
A_{\rm g}\exp\!\left[-\frac{(R-R_{\rm gap})^2}{2\sigma_{\rm g}^2}\right],
\end{aligned}
\label{eq:supp_mock_sigma}
\end{equation}
with negative values clipped to zero and the final profile normalized to
unit maximum.  In the Monte Carlo stack, the ring and gap widths were
tied to the rescaled shell width:
\begin{equation}
\begin{aligned}
\sigma_{\rm r}=\mathrm{clip}(0.8\,\Delta_{\rm phys},0.6,2.0)\,{\rm kpc},
\\
\sigma_{\rm g}=\mathrm{clip}(1.2\,\Delta_{\rm phys},0.8,3.0)\,{\rm kpc}.
\end{aligned}
\end{equation}
The stacked radial profile and corresponding axisymmetric projected
image are shown in Figs.~\ref{fig:supp_mock_profile} and
\ref{fig:supp_mock_projection}.

\section{Formation interpretation and outlook.}
The advantage of the present construction is not that it replaces
hydrodynamical or cosmological formation modeling.  Its role is to
isolate the radial selection step required once an old compact core and
a later settled gas reservoir are present.  A plausible history is the
following.  A main halo forms in a relatively low-density environment
while remaining connected to a cold-gas supply channel
\citep{Sancisi2008,Keres2005,Dekel2009}.  Early
low-angular-momentum gas falls deeper into the halo, cools efficiently,
and forms the old central spheroidal component.  Mildly anisotropic
early accretion, gentle relaxation, or weak mergers may leave this core
slightly triaxial without producing a strongly barred disk.

Later, higher-angular-momentum gas from the same large-scale supply can
enter the halo directly or through gas-rich, star-poor subhalos.  Such
subhalos may remain optically dark if their potential wells are shallow,
their gas surface densities are low, or the ultraviolet background
suppresses star formation \citep{Efstathiou1992}.  If they enter the main
halo on large impact parameter orbits, their H~{\sc i} can be stripped
without leaving an obvious stellar tidal tail.  The stripped gas inherits the orbital
angular momentum of the infall, dissipates radial energy, and partially
circularizes into an extended finite-radius reservoir.  Continued supply
from the same channel can maintain an H~{\sc i} structure larger than
the optical ring and can produce a mild warp if the later angular-
momentum distribution is broader than that of the initial ring-forming
event.

In this interpretation, the Gaussian shell in Eq.~(\ref{eq:phi}) is the
lowest-order radial imprint of the settled H~{\sc i} reservoir.  It is
not a literal self-consistent gravitational potential of the gas.  A
weak quadrupole from a mildly triaxial central body may modulate the
reservoir density and help selected annuli cross a star-formation
threshold, while remaining too weak to destroy the near-circular outer
ring.  The boundary-supported max--min branch supplies the missing
selection step: it separates the outer settling basin from the compact
inner component and keeps the intermediate region underpopulated.  This
is why the absence of an interior $x_{\rm core}$ is not a defect of the
model; the core is a pre-existing boundary state, while the shell
organizes only the newly accreted material outside it.

The mechanism is therefore complementary to both collisional-wave and
bar-resonance pictures.  It does not require a recent intruder launching
an expanding density wave, nor a strong long-lived bar.  It also differs
from a fully self-gravitating massive-ring orbital model: the shell is
an effective localized control deformation associated with settled
accreted material.  The immediate tests are correspondingly direct.
Clean Hoag-type systems should show an old central population, a younger
or actively star-forming outer ring, a gap radius correlated with the
H~{\sc i} shell or reservoir radius, and a ring with small radial motion
relative to its azimuthal motion.  Environmentally processed systems,
such as JO171, need not lie on the same clean shell-localized scaling
relation.

Several extensions are direct.  A multi-shell control potential,
\begin{equation}
\Phi(x)=-\frac{\mu}{x}+
\sum_i\epsilon_i
\exp\!\left[-\frac{(x-x_{0,i})^2}{2\Delta_i^2}\right],
\end{equation}
would test whether double-ring systems such as PGC 1000714 can be
understood as sequential accretion-supported shells.  A dynamical
follow-up should integrate ensembles of dissipative gas parcels in the
time-dependent shell potential, include angular-momentum dispersion and
shell growth, and measure whether
\begin{equation}
\frac{\Sigma(x_{\rm r})}{\Sigma(x_{\rm g})}>C_{\rm gap},
\qquad
\frac{\Delta x_{\rm ring}}{x_{\rm r}}<q_{\rm ring},
\qquad
\frac{\langle |v_R|\rangle_{\rm ring}}
{\langle v_\varphi\rangle_{\rm ring}}<\eta .
\end{equation}
These diagnostics would distinguish a dissipatively settled ring from an
expanding collisional wave and would connect the boundary-supported
control mechanism to observable H~{\sc i}, stellar-age, and kinematic
maps.

\subsection{{Limitations and Multi-Shell Extension}}

The control model is deliberately minimal.  First, the Gaussian shell is
an effective reduced deformation, not a self-consistent density
distribution.  Second, extrema of $W(x)$ identify barrier-like and
well-like radial zones, but they do not define a stellar distribution
function.  Third, the Monte Carlo scan measures parameter-space
occupancy under specified priors; it is not a Bayesian fit to a galaxy.
Fourth, the observational sample is small and heterogeneous, so the
comparison should be interpreted as a morphology-scale consistency test.
Fifth, the velocity-normalized $j_{\rm phys}$ values use heterogeneous
kinematic normalizations and are not recovered galaxy angular-momentum
distributions.

This form is natural for systems with more than one annular component,
such as PGC 1000714 \citep{MutluPakdil2017PGC1000714}.  The criterion
should then be applied locally: for each shell, one searches for a
nearby barrier--well pair with $x_{{\rm g},i}\simeq x_{0,i}$ and
$x_{{\rm r},i}>x_{0,i}$.  If shells overlap strongly, the local
classification may fail and the full ordered set of extrema of $W(x)$
must be analyzed.

A self-consistent next step would replace the radial proxy by a
distribution-function or orbit-based model \citep{BinneyTremaine2008},
project the resulting density field, and fit surface-brightness,
H~{\sc i}, stellar-age, and kinematic data simultaneously.

\section*{Acknowledgements}

This work was supported in part by the National Natural Science Foundation
of China (62505007), and in part by the Boya Postdoctoral Fellowship of
Peking University.  We thank the editors and the anonymous referee for their
constructive comments, which improved the clarity and presentation of this
work.  The authors declare no conflicts of interest.

\section*{Data Availability}
The observational quantities used in this work are taken from the cited
literature. The numerical catalogues and scripts generated for this
study are available in the public GitHub reproducibility repository:
\url{https://github.com/Scarlettxy3/hoag-boundary-layering}.

\bibliographystyle{mnras}
\bibliography{refs}

@article{LyndsToomre1976,
  author  = {Lynds, R. and Toomre, A.},
  title   = {On the interpretation of ring galaxies: the binary ring system II Hz 4},
  journal = {Astrophysical Journal},
  year    = {1976},
  volume  = {209},
  pages   = {382--388},
  doi     = {10.1086/154730}
}

@article{AppletonStruckMarcell1996,
  author  = {Appleton, P. N. and Struck-Marcell, C.},
  title   = {Collisional ring galaxies},
  journal = {Fundamentals of Cosmic Physics},
  year    = {1996},
  volume  = {16},
  pages   = {111--220}
}

@article{ButaCombes1996,
  author  = {Buta, R. and Combes, F.},
  title   = {Galactic rings},
  journal = {Fundamentals of Cosmic Physics},
  year    = {1996},
  volume  = {17},
  pages   = {95--281}
}

@article{Schweizer1987,
  author  = {Schweizer, F. and Ford, W. K. and Jedrzejewski, R. and Giovanelli, R.},
  title   = {The structure and evolution of Hoag's object},
  journal = {Astrophysical Journal},
  year    = {1987},
  volume  = {320},
  pages   = {454--463},
  doi     = {10.1086/165562}
}

@article{Finkelman2011Hoag,
  author  = {Finkelman, I. and Moiseev, A. and Brosch, N. and Katkov, I.},
  title   = {Hoag's Object: evidence for cold accretion on to an elliptical galaxy},
  journal = {Monthly Notices of the Royal Astronomical Society},
  year    = {2011},
  volume  = {418},
  number  = {3},
  pages   = {1834--1849},
  doi     = {10.1111/j.1365-2966.2011.19601.x}
}

@article{FinkelmanBrosch2011UGC4599,
  author  = {Finkelman, I. and Brosch, N.},
  title   = {UGC 4599: a photometric study of the nearest Hoag-type ring galaxy},
  journal = {Monthly Notices of the Royal Astronomical Society},
  year    = {2011},
  volume  = {413},
  number  = {4},
  pages   = {2621--2632},
  doi     = {10.1111/j.1365-2966.2011.18330.x}
}

@article{MutluPakdil2017PGC1000714,
  author  = {Pakdil, Burcin Mutlu and Mangedarage, Mithila and Seigar, Marc S. and Treuthardt, Patrick},
  title   = {A photometric study of the peculiar and potentially double ringed, non-barred galaxy: PGC 1000714},
  journal = {Monthly Notices of the Royal Astronomical Society},
  year    = {2017},
  volume  = {466},
  number  = {1},
  pages   = {355--368},
  doi     = {10.1093/mnras/stw3107}
}

@article{Moretti2018JO171,
  author  = {Moretti, A. and Poggianti, B. M. and Gullieuszik, M. and Mapelli, M. and Jaffe, Y. L. and Fritz, J. and Biviano, A. and Fasano, G. and Bettoni, D. and Vulcani, B. and D'Onofrio, M.},
  title   = {GASP. V. Ram-pressure stripping of a ring Hoag's-like galaxy in a massive cluster},
  journal = {Monthly Notices of the Royal Astronomical Society},
  year    = {2018},
  volume  = {475},
  number  = {3},
  pages   = {4055--4065},
  doi     = {10.1093/mnras/sty085}
}

@article{Bannikova2018,
  author  = {Bannikova, Elena Yu.},
  title   = {The structure and stability of orbits in Hoag-like ring systems},
  journal = {Monthly Notices of the Royal Astronomical Society},
  year    = {2018},
  volume  = {476},
  number  = {3},
  pages   = {3269--3277},
  doi     = {10.1093/mnras/sty444}
}

@book{BinneyTremaine2008,
  author    = {Binney, James and Tremaine, Scott},
  title     = {Galactic Dynamics},
  edition   = {2},
  publisher = {Princeton University Press},
  address   = {Princeton, NJ},
  year      = {2008},
  doi       = {10.1515/9781400828722}
}

@book{Strogatz2015,
  author    = {Strogatz, Steven H.},
  title     = {Nonlinear Dynamics and Chaos: With Applications to Physics, Biology, Chemistry, and Engineering},
  edition   = {2},
  publisher = {Westview Press},
  address   = {Boulder, CO},
  year      = {2015},
  isbn      = {9780813349107}
}

@book{Arnold1986,
  author    = {Arnold, V. I.},
  title     = {Catastrophe Theory},
  edition   = {2},
  publisher = {Springer},
  address   = {Berlin},
  year      = {1986},
  doi       = {10.1007/978-3-642-96937-9}
}

@article{Sancisi2008,
  author  = {Sancisi, R. and Fraternali, F. and Oosterloo, T. and van der Hulst, T.},
  title   = {Cold gas accretion in galaxies},
  journal = {Astronomy and Astrophysics Review},
  year    = {2008},
  volume  = {15},
  pages   = {189--223},
  doi     = {10.1007/s00159-008-0010-0}
}

@article{Keres2005,
  author  = {Kere{\v{s}}, D. and Katz, N. and Weinberg, D. H. and Dav{\'e}, R.},
  title   = {How do galaxies get their gas?},
  journal = {Monthly Notices of the Royal Astronomical Society},
  year    = {2005},
  volume  = {363},
  number  = {1},
  pages   = {2--28},
  doi     = {10.1111/j.1365-2966.2005.09451.x}
}

@article{Dekel2009,
  author  = {Dekel, A. and Birnboim, Y. and Engel, G. and Freundlich, J. and Goerdt, T. and Mumcuoglu, M. and Neistein, E. and Pichon, C. and Teyssier, R. and Zinger, E.},
  title   = {Cold streams in early massive hot haloes as the main mode of galaxy formation},
  journal = {Nature},
  year    = {2009},
  volume  = {457},
  pages   = {451--454},
  doi     = {10.1038/nature07648}
}

@article{Efstathiou1992,
  author  = {Efstathiou, G.},
  title   = {Suppressing the formation of dwarf galaxies via photoionization},
  journal = {Monthly Notices of the Royal Astronomical Society},
  year    = {1992},
  volume  = {256},
  pages   = {43P--47P},
  doi     = {10.1093/mnras/256.1.43P}
}

\label{lastpage}

\end{document}